\documentstyle[11pt,newpasp,twoside,epsf]{article}
\def\edcomment#1{\iffalse\marginpar{\raggedright\sl#1\/}\else\relax\fi}
\reversemarginpar

\begin{document}
\title{GAIA Photometric Performances}
\author{Vladas Vansevi\v{c}ius}
\affil{Institute of Physics, Go\v{s}tauto 12, Vilnius 2600, Lithuania,
wladas@astro.lt}
\author{Audrius Brid\v{z}ius}
\affil{Institute of Physics, Go\v{s}tauto 12, Vilnius 2600, Lithuania,
bridzius@astro.lt}

\begin{abstract}
GAIA present day (end-2002) photometric performances, estimated basing on the
stellar populations crucial for understanding the formation and evolution of
the Galaxy, are discussed. Performance of the GAIA photometric systems (PSs) is
evaluated taking into account their ability to simultaneously determine the
main stellar parameters: $T_{\rm eff}$, $\log\,g$, [M/H], and $E_{B-V}$, for a
large variety of stars down to $V\sim20$. A sample of the stars (Photometric
System Test Targets, PSTTs) applicable for evaluation of the GAIA photometric
systems is presented. Definitions of the 1X PS and its accuracies of stellar
parameterization are given. We conclude that there is still no photometric
system, which would allow to achieve the scientific objectives of the GAIA
mission at the limiting magnitude $V=20$, proposed to date.
\end{abstract}

\section{Introduction}

Though the primary goal of the GAIA mission is a study of the Galactic
dynamics, photometric as well as spectroscopic observations are also of great
importance in order to achieve the main GAIA objectives (ESA 2000). Therefore,
an optimal multicolour photometric system, adequate to the main mission goals,
has to be designed and analysed in detail. One of the most important steps
towards such a PS is an elaboration of a method for unbiased evaluation of the
PSs proposed for GAIA in terms of the accuracy of stellar parameters derived
(Vansevi\v{c}ius, Brid\v{z}ius, \& Drazdys 2002). A tight time schedule of the
pre-launch preparation constrains possibilities of thorough testing of the new
PS on real sky objects. Therefore, optimization and further choice of the
most suitable PS for the GAIA mission must be entirely based on the synthetic
and observed spectral energy distributions (SEDs) and the representative set of
a large variety of stellar populations and interstellar medium inside the
Galaxy.

We would like to note two key points which have been taken into account for
evaluating the GAIA PSs in the present study. The same GAIA observation
(Deveikis, Brid\v{z}ius, \& Vansevi\v{c}ius 2002) and parameterization
(Brid\v{z}ius \& Vansevi\v{c}ius 2002) methods have been applied to all PSs,
and the performance of the PSs has been tested on the PSTT stars, which
are of the highest importance to the GAIA mission (ESA 2000).

\section{Requirements for the GAIA Photometric System}

The primary objective of the GAIA mission is to obtain data which allows us to
study the composition, formation and evolution of the Galaxy. This priority also
constrains the requirements for the GAIA PS. Therefore, metallicity and age
variations (RGB, AGB and HB), star formation history (SFH) over the last 14 Gyr
(A-K MS stars), and dust distribution within the Galaxy are the most important
issues to be assessed by the GAIA PS.

The variety of the object types and astrophysical circumstances presuppose
specific requirements for the GAIA PS performance. It is indisputable that the four
main parameters ($T_{\rm eff}$, $\log\,g$, [M/H], and $E_{B-V}$) – ought to be
derived as precisely as possible at the limiting magnitude of the survey, $V=20$.
However, it is very important to decide which parameters should be determined
additionally. Therefore, below we list the most relevant items whose various degree
of treatment could significantly influence the final design of the GAIA PS:
\begin{itemize}
\item variation of [$\alpha$/Fe] - should it be determined by PS?
\item extinction of the F-K type stars - should it be determined for each individual
star by PS or assigned from the 3-D extinction maps?
\item binaries - should PS be adapted for their recognition and parameter estimation?
\item variable extinction law, $R_{BV}$ - should it be determined by PS?
\item chemical peculiarities - should the strong peculiar features (e.g. C, N) be
avoided or measured by PS?
\item WD, BD, WR, $\delta$Cep, T Tau, Mira, C, etc. type stars - which parameters
should be derived by PS?
\end{itemize}

Although it is very important to take all mentioned points into account, present
day investigations and discussions are devoted mostly to the first two items.
Determination of the interstellar extinction of the F-K type stars is fully
accounted (S\={u}d\v{z}ius et al. 2002) and can be discussed quantitatively.
However, the [$\alpha$/Fe] problem is still pending due to lack of a proper grid
of SEDs. Therefore, only these two items are taken into account for selection
of the PSTTs.

\section{GAIA Photometric System Test Targets (PSTTs)}

According to ESA (2000), the stars to be observed with GAIA down to $G\sim20$
can be roughly represented by the following proportions of stellar populations
(Ms - stands for millions of stars):
\begin{itemize}
\item the total number of stars of all types $\ga1000$ Ms;
\item disk dwarfs $\sim780$ Ms; disk giants $\sim90$ Ms;
\item thick disk (all types) $\sim100$ Ms; halo (all types) $\sim70$ Ms;
\end{itemize}
and spectral types:
\begin{itemize}
\item O-B $\sim2$ Ms; A $\sim50$ Ms; F $\sim230$ Ms;
\item G $\sim400$ Ms; K $\sim300$ Ms; M $\sim55$ Ms.
\end{itemize}

In order to evaluate any PS proposed for GAIA we need to select some set of
standard objects with which to test a PS. PSTTs ought to cover the most
important stellar population types mentioned above and the most
characteristic variety of astrophysical circumstances expected to be
encountered in the Galaxy. A set of stellar types which would make up a minimum
number of PSTTs is presented in Table 1 and in Figure 1. We need to keep this list
as short as possible in order to quickly test any new improvements of the GAIA
PSs. However, it should also be comprehensive in the sense of the main GAIA
goals. The same statements are also valid for other parameters listed below.
The proper choice of the PSTTs is a very important issue for GAIA PS
development and optimization, therefore, the list is open for discussions
and suggestions.

\begin{figure}[t]
\plotone{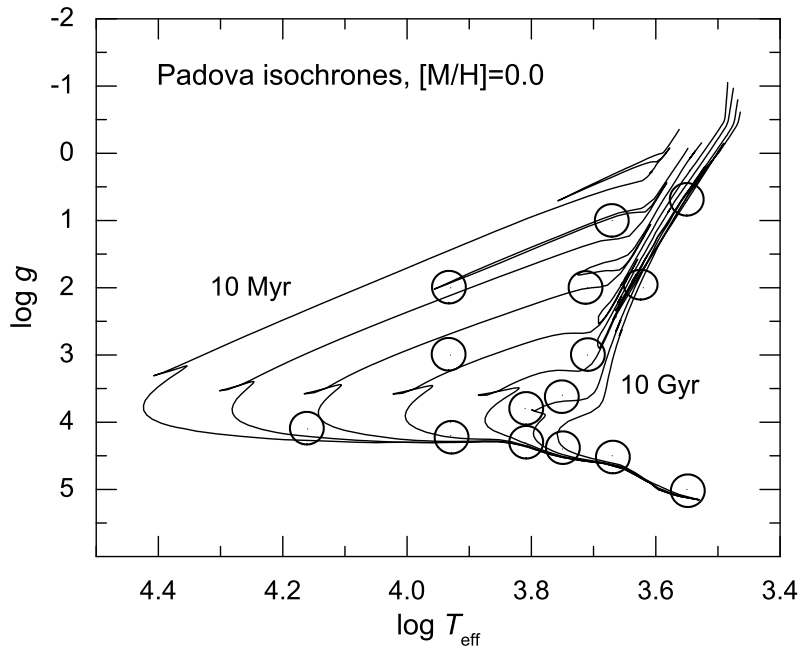}
\caption{GAIA Photometric System Test Targets (PSTTs).}
\end{figure}

\begin{table}
\caption{GAIA Photometric System Test Targets (PSTTs).}
\begin{center}
\begin{tabular}{cccccc}
\tableline
Code & Sp. Type & $T_{\rm eff}$ & $\log\,T_{\rm eff}$ & $\log\,g$ & $M_V$ \\
\tableline
 1 & B V   & 14500 &  4.16 &  4.1 & -0.9 \\
 2 & A V   &  8500 &  3.93 &  4.2 &  1.6 \\
 3 & F V   &  6450 &  3.81 &  4.3 &  3.7 \\
 4 & G V   &  5600 &  3.75 &  4.4 &  5.1 \\
 5 & K V   &  4700 &  3.67 &  4.5 &  6.7 \\
 6 & M V   &  3550 &  3.55 &  4.7 & 10.3 \\
 7 & F IV  &  6450 &  3.81 &  3.8 &  2.6 \\
 8 & G IV  &  5600 &  3.75 &  3.6 &  3.1 \\
 9 & G III &  5150 &  3.71 &  3.0 &  0.9 \\
10 & K III &  4150 &  3.62 &  2.5 &  0.3 \\
11 & M III &  3550 &  3.55 &  1.2 & -0.5 \\
12 & BHB   &  8500 &  3.93 &  3.0 &  0.5 \\
13 & RHB   &  5150 &  3.71 &  2.0 &  0.5 \\
14 & A I   &  8500 &  3.93 &  2.0 & -5.0 \\
15 & G I   &  4700 &  3.67 &  1.0 & -4.5 \\
\tableline
\end{tabular}
\end{center}
\end{table}

The list of PSTTs is complemented by other astrophysical parameters,
representing circumstances which could be encountered in the Galaxy
(Table 2), as completely as possible. Two variants of a complete set of
models based on distance, D, of the PSTTs or their apparent magnitude,
$V_{\rm J}$, are foreseen. In order to make a grid of PSTTs convenient
for computer modelling and analysis, we suggest to use a complete set of
various combinations even if one supposes that any particular combination
of the parameters is unrealistic and cannot be found in the Galaxy.
Therefore, the complete set of models should contain 5625 different
combinations of the parameters in total. Such a task can be easily
solved even on a PC type computer.

\begin{table}
\caption{Parameters for the PSTTs. $V_{\rm J}$ -- $UBV$ Johnson system.}
\begin{center}
\begin{tabular}{ccccccc}
\tableline
No.  & Parameter &  &  &  &  & \\
\tableline
1 & D, kpc        &  1   & 2   & 4    & 8    & 20    \\
2 & $V_{\rm J}$   &  15  & 17  & 18   & 19   & 20    \\
3 & $E_{B-V}$     & 0.05 & 0.5 & 1.0  & 1.5  & 3.0   \\
4 & [M/H]         & +0.5 & 0.0 & -0.5 & -1.5 & -2.5  \\
5 & [$\alpha$/Fe] & 0.0  & 0.3 & 0.6  & --   & --    \\
\tableline
\end{tabular}
\end{center}
\end{table}

\section{The 1X GAIA Photometric System}

Few different photometric systems have been proposed for the GAIA mission to
date: 1F (ESA 2000) and its modification 2F, 2A (Munari 1999), 2G and its
modification 3G (H{\o}g, Strai\v{z}ys, \& Vansevi\v{c}ius 2000). However,
the analysis of their performance has demonstrated that none of them satisfy the
requirements for the GAIA PS (Vansevi\v{c}ius et al. 2002). Therefore, taking
all valuable findings in those PSs into account, a new photometric system,
1X, is proposed. Evolution of the GAIA PSs and the limiting magnitudes at which they
satisfy GAIA goals are shown in Table 3.

\begin{table}
\caption{Evolution of the GAIA PSs. $V_{\rm J}$ indicates the limiting magnitude
at which PS performance is satisfactory for the GAIA goals.}
\begin{center}
\begin{tabular}{ccccc}
\tableline
Proposed, year & & Photometric System  & & Limit, $V_{\rm J}$ \\
\tableline
1999 & \bf{1F} & \bf{2A}  & \bf{2G} & 17  \\
2000 &         &          & \bf{3G} & 18  \\
2001 & \bf{2F} &          &         & 18  \\
\tableline
2002 &         & \bf{1X}  &         & 19  \\
2003 &         &   ???    &         & 20  \\
\tableline
\end{tabular}
\end{center}
\end{table}

The filter transmission curves of the 1X medium band (MB) PS designed for the
Spectro telescope are shown in Figure 2. The central wavelength and width of
the bands as well as a number of slots, allocated for each filter in the focal
plane, are given in Table 4.

\begin{table}
\caption{Definition of the 1X PS. CWL - central wavelength; FWHM - full width
of the filter transmission band at half maximum; NF - number of slots allocated
to the filter in the focal plane.}
\begin{center}
\begin{tabular}{cccccccccccc}
\tableline
Filter & x33 & x38 & x41 & x46 & x51 & x55 & x65 & x78 & x82 & x86 & x99 \\
\tableline
CWL    & 323 & 382 & 410 & 465 & 510 & 555 & 655 & 785 & 825 & 860 & 992 \\
FWHM   & 63  & 29  & 19  & 29  & 19  & 49  & 29  & 29  & 29  & 29  & 56  \\
NF     &  2  &  2  &  2  &  1  &  2  &  1  &  1  &  1  &  1  &  1  &  1  \\
\tableline
\end{tabular}
\end{center}
\end{table}

\begin{figure}[t]
\plotfiddle{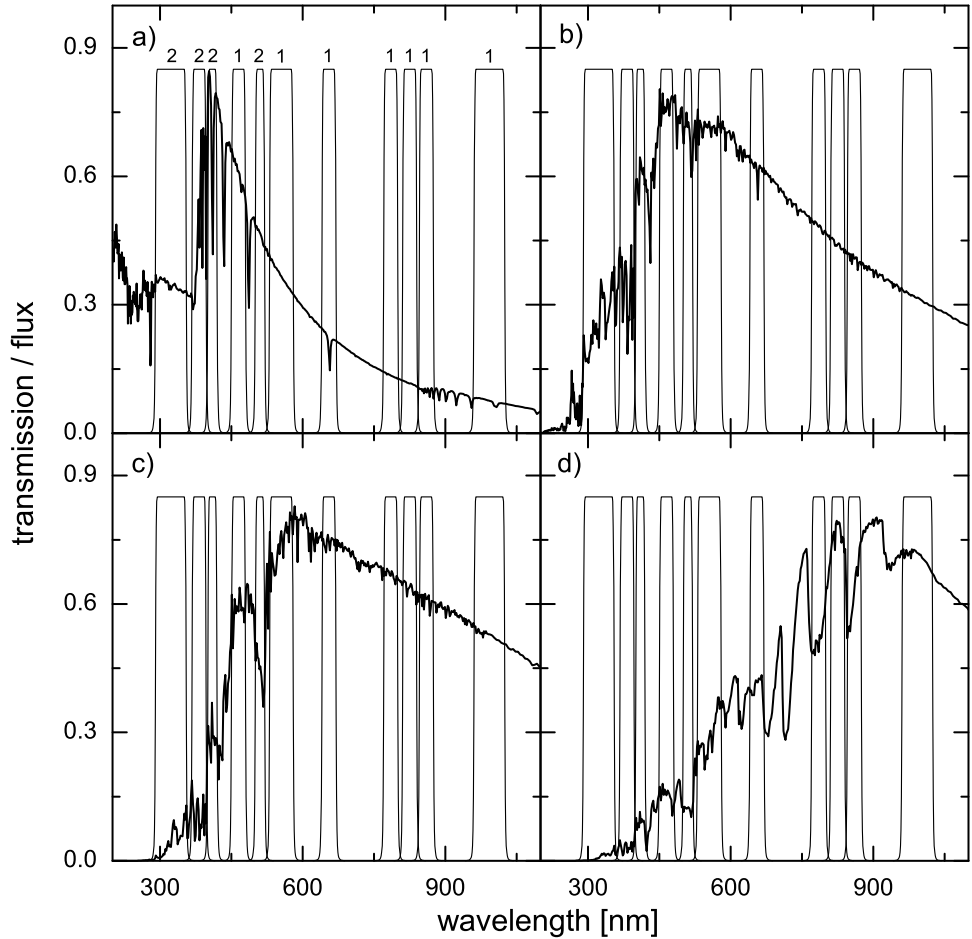}{12cm}{0}{70}{70}{-180}{-10}
\caption{The 1X GAIA PS. SEDs of the main sequence stars of $T_{\rm eff}$:
a) 9000 K, b) 5500 K, c) 4500 K, d) 3500 K, are plotted. The numbers of
slots allocated to each filter in the focal plane are shown in panel a.}
\end{figure}

\section{Performance of the 1X Photometric System}

The performance of the 1X PS is evaluated in terms of its ability to derive basic
stellar parameters: $T_{\rm eff}$, $\log\,g$, [M/H], and $E_{B-V}$, for the
PSTTs. The procedure for simulation of the photometric GAIA observations developed
by Deveikis et al. (2002) was used. The simultaneous 4-D stellar parameterization
method (Brid\v{z}ius \& Vansevi\v{c}ius 2002) was applied.

Some specific assumptions on the 1X PS application to the GAIA case introduced in
the modelling procedure should be noted:
\begin{itemize}
\item the design of GAIA-1 is assumed as it is given in ESA (2000);
\item only the medium band PS is tested assuming that there is no supplementary
information obtained from the GAIA broad band (BB) PS (negligible influence of
the BB PS to the performance of MB PS was demonstrated by Vansevi\v{c}ius et
al. 2002);
\item the total number of slots for MBP filters is assumed to be equal to 15
(as it is foreseen in the Radial Velocity Spectrometer (RVS) design for GAIA-2);
\item the maximum of filter transmission curves is set conservatively to 85$\%$;
\item modelling is performed by employing the SEDs from BaSeL 2.2 (Lejeune,
Cuisinier, \& Buser 1998) with solar abundance of $\alpha$-elements,
[$\alpha$/Fe]=0.0;
\item the standard ($R_{BV}=3.1$; Cardelli, Clayton, \& Mathis 1989) and invariable
extinction law is assumed;
\item the aperture photometry procedure is applied for the modelling of star measurement.
\end{itemize}

The results of the 1X PS performance test are presented in Table 5. The typical value
of colour excess, $E_{B-V}$, was assumed for each particular PSTT star. All apparent
magnitudes were set to $V=18$, therefore, each PSTT was placed at some distance, D.
In order to derive the accuracy of the 1X PS more reliably, 100 independent simulations
of each PSTT were performed, and deviations of the stellar parameters from the true
value were determined. The r.m.s. estimate was employed, and the $\sigma$'s of the
stellar parameters are tabulated.

\begin{table}
\caption{Performance of the 1X GAIA PS evaluated on PSTTs of solar abundance,
[M/H]=0.0, at $V=18$. SpT -- spectral type.}
\begin{center}
\begin{tabular}{ccccccccc}
\tableline
Code & SpT & $M_V$ & $E_{B-V}$ & D,kpc & $\sigma T_{\rm eff}$,\%
& $\sigma$($\log\,g$) & $\sigma$[M/H] & $\sigma A_V$      \\
\tableline
1  & B V   & -0.9 & 3    & 0.8 & 4.3 & 0.10 & 1.20 & 0.02 \\
2  & A V   & 1.6  & 1    & 4.6 & 4.0 & 0.15 & 0.30 & 0.09 \\
3  & F V   & 3.7  & 1    & 1.7 & 2.3 & 0.35 & 0.05 & 0.09 \\
4  & G V   & 5.1  & 0.5  & 1.9 & 2.8 & 0.30 & 0.20 & 0.10 \\
5  & K V   & 6.7  & 0.5  & 0.9 & 1.5 & 0.30 & 0.05 & 0.09 \\
6  & M V   & 10.3 & 0.05 & 0.3 & 0.5 & 0.10 & 0.05 & 0.05 \\
7  & F IV  & 2.6  & 1    & 2.9 & 2.8 & 0.25 & 0.05 & 0.10 \\
8  & G IV  & 3.1  & 1    & 2.3 & 3.3 & 0.50 & 0.25 & 0.11 \\
9  & G III & 0.9  & 1    & 6.3 & 1.0 & 0.45 & 0.05 & 0.03 \\
10 & K III & 0.3  & 1    & 8.3 & 0.5 & 0.05 & 0.05 & 0.04 \\
11 & M III & -0.5 & 1    & 12. & 0.5 & 0.15 & 0.10 & 0.09 \\
12 & BHB   & 0.5  & 1    & 7.6 & 2.0 & 0.10 & 0.60 & 0.04 \\
13 & RHB   & 0.5  & 1    & 7.6 & 2.0 & 0.55 & 0.15 & 0.08 \\
14 & A I   & -5.0 & 3    & 5.5 & 0.8 & 0.05 & 0.10 & 0.01 \\
15 & G I   & -4.5 & 3    & 4.4 & 1.0 & 0.15 & 0.05 & 0.06 \\
\tableline
\end{tabular}
\end{center}
\end{table}

We find that performance of the 1X PS at $V=18$ is excellent and completely
satisfies the requirements raised for the GAIA PS by the main GAIA objectives.
The performance of the 1X PS is significantly better, especially if cases of
higher extinction are considered, comparing to the previous PSs, 2F \& 3G
(Figures 3 -- 5; $V=18$; the r.m.s. estimates of the parameters were derived
basing on 100 simulations of each PSTT; for more details see Vansevi\v{c}ius
et al. 2002). The 1X PS also performs satisfactory at $V=19$, however, there
is an obvious necessity to make it more accurate at $V=20$.

\begin{figure}[t]
\plotfiddle{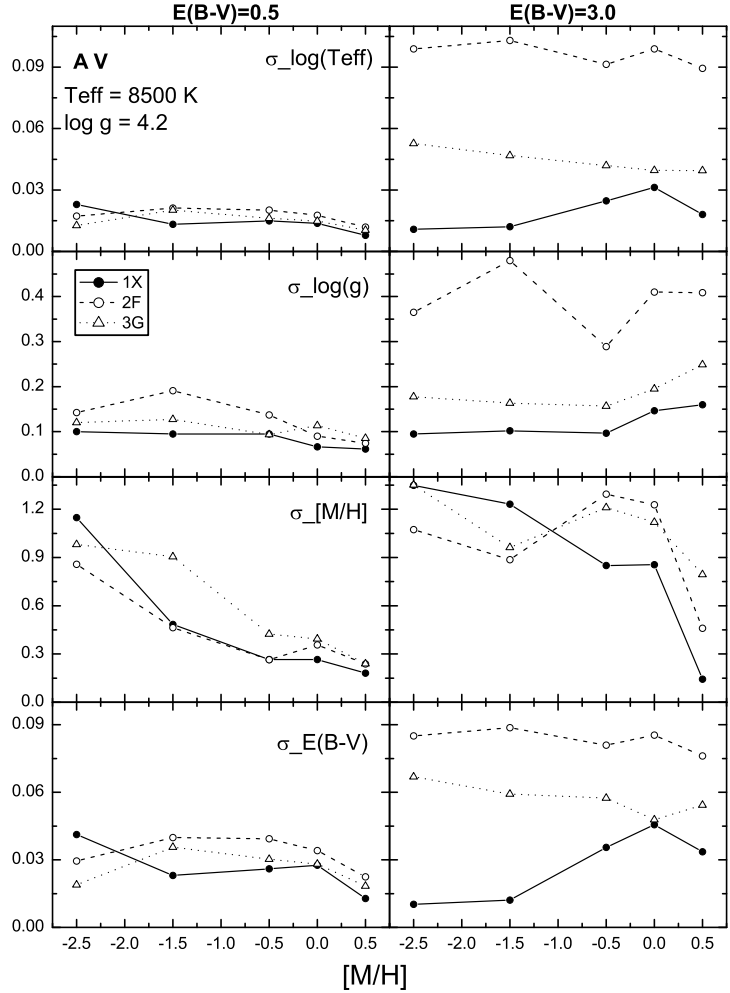}{13cm}{0}{63}{63}{-170}{-45}
\caption{Comparison of the performance of the 1X, 2F, \& 3G GAIA PSs, PSTT \#2, see Table 1.}
\end{figure}

\begin{figure}[t]
\plotfiddle{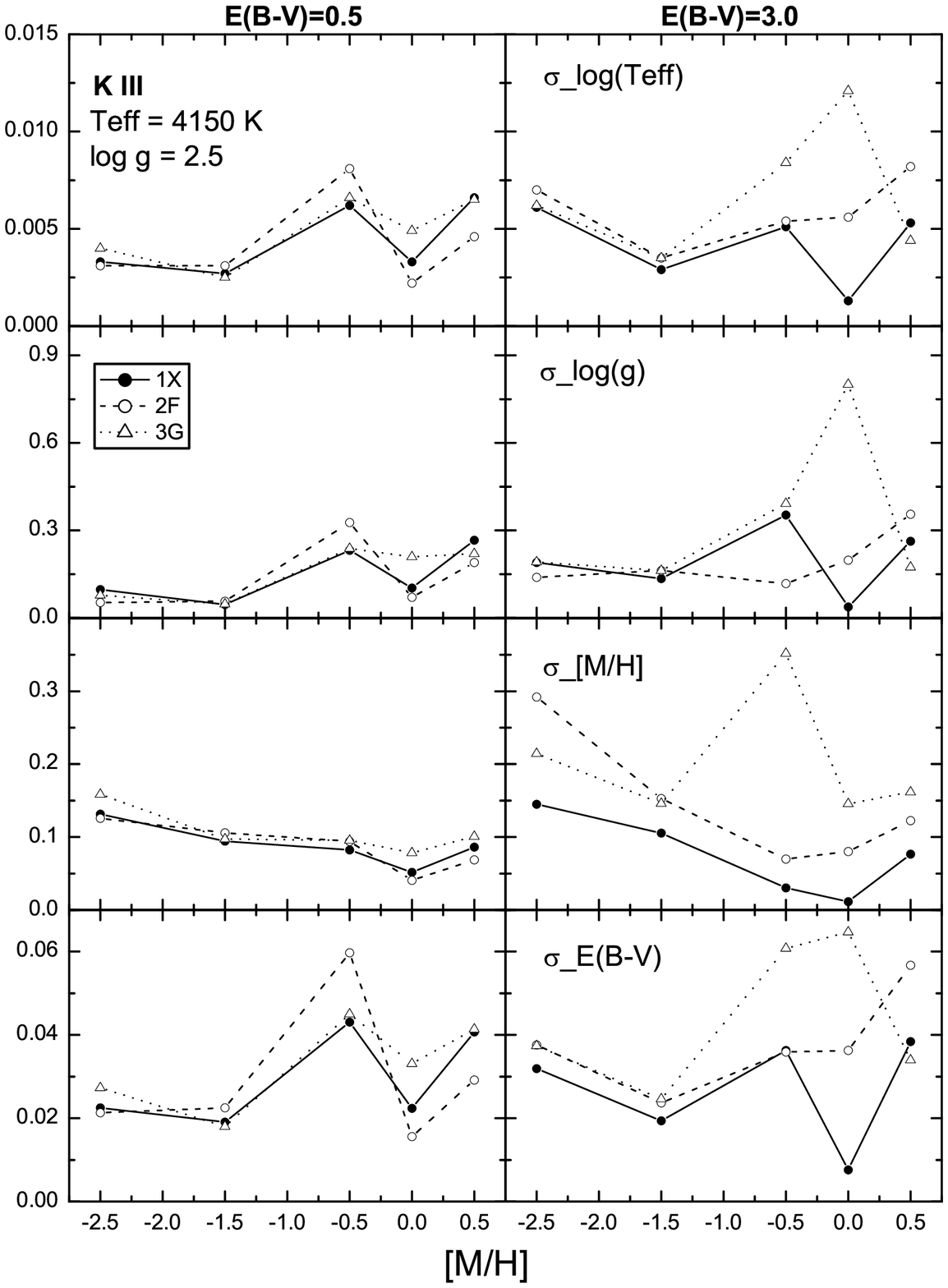}{13cm}{0}{63}{63}{-170}{-45}
\caption{Comparison of the performance of the 1X, 2F, \& 3G GAIA PSs, PSTT \#10, see Table 1.}
\end{figure}

\begin{figure}[t]
\plotfiddle{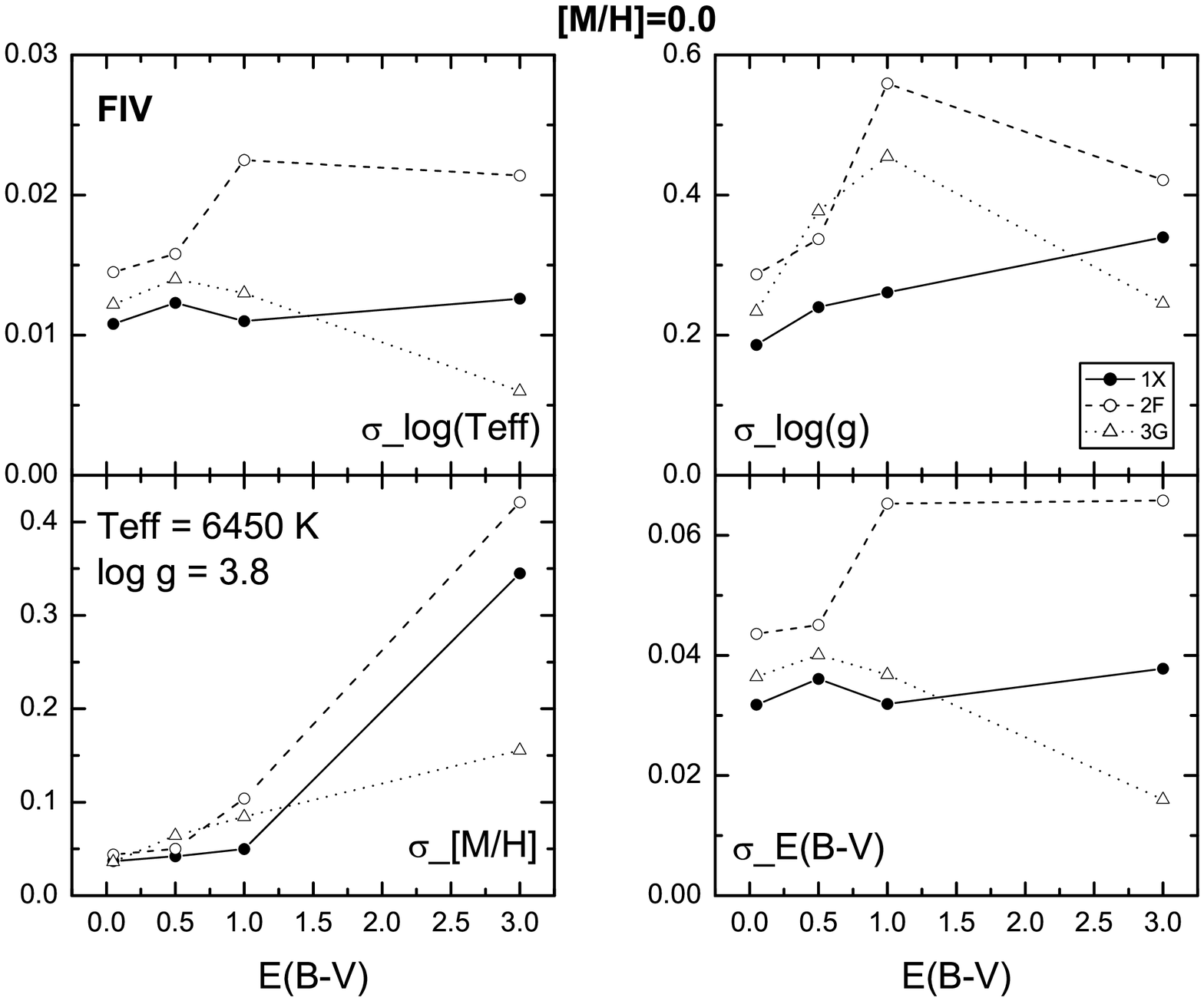}{9cm}{0}{55}{55}{-200}{-20}
\caption{Comparison of the performance of the 1X, 2F, \& 3G GAIA PSs, PSTT \#7, see Table 1.}
\end{figure}

\section{Conclusions}

We have suggested a set of the GAIA Photometry System Test Targets (PSTTs)
selected in order to facilitate easy and comprehensive test of the proposed
GAIA PSs. We have introduced a new 1X PS which demonstrates superior performance
compared to the PSs proposed for GAIA earlier. The performance of the 1X PS
was evaluated taking into account its capability to simultaneously determine the
main stellar parameters: $T_{\rm eff}$, $\log\,g$, [M/H], and $E_{B-V}$, for the
PSTT stars, and assuming that no supplementary information is available, except
the data obtained by the GAIA medium band PS.

GAIA is assumed to measure stars in the Galaxy down to $V\sim20$, however, even
the best photometric system (1X) among the PSs, proposed for GAIA to date (end-2002),
performs satisfactory only down to $V\sim19$. Therefore, we conclude that there is
still no optimal PS, in terms of the main GAIA mission goals (ESA 2000),
proposed to date. However, excellent accuracy of the stellar parameters determined
at $V=18$ (1X PS) implies fulfilment of the requirements for GAIA PS at $V=20$ after
appropriate fine tuning of each band.

Some gain in performance could be achieved by applying different parameterization
schemes for the faintest stars ($V>18$). Parameterization of these stars should be
performed by employing all complementary information obtained by GAIA (parallaxes,
spectra, 3-D extinction maps constructed basing on $E_{B-V}$ derived for the brighter
stars, $V<18$, etc.). However, such an automatic parameterization procedure is not
yet developed.

In order to optimize the GAIA PS, the following supplemental information is urgently
needed:
\begin{itemize}
\item a homogeneous and complete database of theoretical stellar spectra, especially
SEDs of various [$\alpha$/Fe];
\item realistic estimates of the limiting magnitudes and accuracies of the stellar
parameters derived from the RVS data;
\item determination of the limiting distance, resolution and achievable accuracy
of the 3-D extinction maps of the Galaxy;
\item realistic models of the photometry procedure, taking into account real
sky complexity and accuracy of the post-mission calibrations.
\end{itemize}

\acknowledgments

This work was supported by a Grant of the Lithuanian State Science and Studies
Foundation. We are indebted to Valdas Vansevi\v{c}ius for his help on preparation
of the article.


\end{document}